\documentclass[prl,aps,floats,twocolumn,prd,showpacs,nofootinbib]{revtex4}

\usepackage{amssymb}
\usepackage{amsmath}

\usepackage{graphicx}
\usepackage{graphics}
\usepackage{dcolumn}
\usepackage{color}
\usepackage{rotate}
\usepackage{fancyhdr}

\def\be{\begin{equation}}
\def\ee{\end{equation}}
\def\bea{\begin{eqnarray}}
\def\eea{\end{eqnarray}}
\def\nn{\nonumber}

\newcommand{\refeq}[1]{Eq.~(\ref{eq:#1})}          
\newcommand{\refeqs}[2]{Eqs.~(\ref{eq:#1})--(\ref{eq:#2})}          
\newcommand{\reffig}[1]{Fig.~\ref{fig:#1}}

\newcommand{\vs}{\nonumber\\}

\newcommand{\ApJ}{Astrophys. J.}

\newcommand{\AsAs}{Astron.  Astrophys.}

\def\wignertj#1#2#3#4#5#6{ \left( \begin{array}{ccc} #1 & #3 & #5
\\ #2 & #4 & #6 \\ \end{array} \right)}

\begin{document}

\title{Rotation of the cosmic microwave background polarization from weak gravitational lensing}

\author{Liang Dai}
\affiliation{Department of Physics and Astronomy, Johns Hopkins University, Baltimore, Maryland 21218, USA}

\date{\today}


\begin{abstract}

When a cosmic microwave background (CMB) photon travels from the surface of last scatter through spacetime metric perturbations, the polarization vector may rotate about its direction of propagation. This gravitational rotation is distinct from, and occurs in addition to, the lensing deflection of the photon trajectory. This rotation can be sourced by linear vector or tensor metric perturbations and is fully coherent with the curl deflection field. Therefore, lensing corrections to the CMB polarization power spectra as well as the temperature-polarization cross-correlations due to non-scalar perturbations are modified. The rotation does not affect lensing by linear scalar perturbations, but needs to be included when calculations go to higher orders. We present complete results for weak lensing of the full-sky CMB power spectra by general linear metric perturbations, taking into account both deflection of the photon trajectory and rotation of the polarization. For the case of lensing by gravitational waves, we show that the $B$ modes induced by the rotation largely cancel those induced by the curl component of deflection.

\end{abstract}



\maketitle


Weak gravitational lensing of the cosmic microwave background (CMB) allows us to probe cosmic structures along the line of sight. Most generally, metric perturbations affects a photon in three ways: the photon energy is shifted; its direction of propagation is deflected; and its plane of polarization is re-oriented. Considerable attention has been focused in the literature on the deflection, mostly by scalar metric perturbations (density perturbations)~\cite{CMBlensing,Hu:2000ee}, an effect that has now been observed~\cite{Smith:2007rg,Hirata:2008cb,Das:2011ak,Keisler:2011aw,Ade:2013tyw}. In particular, deflection induces $B$ modes in the CMB polarization~\cite{Zaldarriaga:1998ar,Seljak:2003pn}, which have recently been detected~\cite{Hanson:2013hsb}. Lensing by vector or tensor metric perturbations---that may arise, for example, from cosmic strings~\cite{Yamauchi:2012bc} or from inflationary gravitational waves~\cite{Mollerach:1997ev,Cooray:2005hm,Li:2006si,Book:2011na}---have also been studied.

Metric perturbations will also rotate the plane of polarization of a photon. In a perturbed spacetime, the polarization vector is parallel-transported along the geodesic, while the reference basis vectors with respect to which a local observer measures the orientation of the polarization are {\it not}. The mismatch results in an observed polarization state that deviates from what one would otherwise observe in flat spacetime. Skrotsky first considered this effect in the context of gravitomagnetic drag due to massive rotating bodies~\cite{Skrotsky:1957}. Later work extended that study to weak gravitational fields generated by other localized masses~\cite{RotationByMasses}. 

However, rotation of polarization has not been properly included in the context of weak lensing of CMB polarization by general non-localized lenses. The effect occurs for vector/tensor linear metric perturbations; it vanishes for scalar perturbations at linear order, but should appear at nonlinear orders. Previously, rotation of polarized radiation from the CMB or from quasars by primordial vector perturbations has been considered~\cite{Morales:2007rd}. Nevertheless, the important implications of the combination of this effect with those of the lensing deflection have not been appreciated. In this Letter, we systematically consider this effect from the most general linear metric perturbations, and find that the rotation angle of the polarization is the same as that for the rotation of the image, thus implying full correlation between the rotation of the polarization and the curl part of the deflection field. Including the rotation, we present complete analytical results for the weak lensing of the CMB temperature and polarization power spectra. We will demonstrate, with weak lensing by gravitational waves, that interference between curl deflection and rotation considerably reduces the $B$-mode power converted from intrinsic $E$-mode power as found previously~\cite{Li:2006si,Rotti:2011aa,Padmanabhan:2013xfa}, making the gradient part and the curl part of the deflection field equally efficient at generating lensing $B$-mode polarizations.

A Friedmann-Robertson-Walker Universe perturbed by the most general linear metric perturbations is described by the following metric,
\bea
ds^2 & = & a^2(\tau) \left[ - (1 + 2 A(\vec x,\tau)) d\tau^2 + 2 B_i(\vec x, \tau) d\tau dx^i \right.\nn\\
&&\left. + \left( \delta_{ij} + h_{ij}(\vec x,\tau) \right) dx^i dx^j \right],
\eea
where $\tau$ is the conformal time and $\vec x$ is the comoving position, and $A$, $B_i$, and $h_{ij}$ parameterize perturbations.

Consider a photon, characterized by its four-momentum $p^{\mu}$ and the polarization vector $\epsilon^{\mu}$, that propagates from the source location to the observer at the origin. Given that the photon is seen by the observer in direction $n^i$ (measured with respect to spatial tetrads $e^{\mu}_{(i)o},\,i=1,2,3$ at observer's location) in the sky, its trajectory is solved from the geodesic equation $dp^{\mu}/d\lambda=-\Gamma^{\mu}_{\alpha\beta}p^{\alpha}p^{\beta}$, with respect to the affine parameter $\lambda$. As the photon travels along the null geodesic, $\epsilon^{\mu}$ is parallel-transported,
\bea
d\epsilon^{\mu}/d\lambda & = & - \Gamma^{\mu}_{\alpha\beta} p^{\alpha} \epsilon^{\beta}.
\eea

At a given location in spacetime, an observer measures photon polarization by projecting $\epsilon^{\mu}$ onto the screen-projected plane perpendicular to both the four-momentum $p^{\mu}$ and the observer's four-velocity~\cite{Pitrou:2008hy}. The polarization can then be expanded using local tetrads with components $\epsilon_{(\nu)}=e^{\mu}_{(\nu)}\epsilon_{\mu}$, which determine the orientation of polarization. A set of local tetrads $e^{\mu}_{(\nu)}$, for $\nu=0,1,2,3$, need to be specified at every spacetime point, since the comparison of the directions of vectors at separated locations is non-trivial in a general perturbed spacetime. Up to linear order in metric perturbations, an irrotational choice can be made,
\bea
\label{eq:tetrad-temporal}
e^{\mu}_{(0)} = a^{-1} \left\{ 1 - A, \vec 0 \right\}, \quad e^{\mu}_{(i)} = a^{-1} \left\{ B_i, \delta^j{}_i - h^j{}_i / 2 \right\},
\eea
which reduces to the natural choice $e^{\mu}_{(\nu)}=a^{-1}\delta^{\mu}_\nu$ in the absence of perturbations. The rotation of polarization due to lensing can be then defined as the unique rotation about the line of sight that takes the observed components $\epsilon_{(i)o}$, for $i=1,2,3$, to what they would be in the {\it absence} of perturbations. Note that the latter would simply be $\epsilon_{(i)s}$ for $i=1,2,3$, at source location, since without perturbations local tetrads would be identical everywhere.

Under the Born approximation, the rotation angle $\psi$ for polarization is obtained by integrating along the {\it unperturbed} line of sight,
\bea
\label{eq:rotation-angle-integral}
\psi(\hat n) = (1/2) \varepsilon^{ij}{}_k n^k \int^{\chi_s}_0 d\chi \left( \partial_i B_j - n^l \partial_i h_{jl} \right).
\eea
Here $\varepsilon_{ijk}$ is the antisymmetric Levi-Civita tensor in three dimensions, and the comoving radial distance $\chi$ from the observer parameterizes the line of sight. Note that a non-zero $\psi$ is not a coordinate artifact. In fact, a gauge transformation modifies \refeq{rotation-angle-integral} by only boundary terms. Those  correspond to Lorentz transformations of the source frame and the observer frame, since \refeq{tetrad-temporal} defines different tetrads in different gauges. 

Unlike Faraday rotation, the rotation due to metric perturbations is achromatic. Scalar metric perturbations, namely the potential $A$, the gradient part of $B_i$, and the trace part and the longitudinal part of $h_{ij}$, do not contribute to $\psi$ at linear order. On the contrary, vector perturbations---the divergence-free part of $B_i$ or the transverse vector parts of $h_{ij}$, as well as tensor perturbations---the transverse-tensor parts of $h_{ij}$, do induce rotation at this order. 

Non-uniform deflection of the photon direction $\theta^i(\hat n)$ by gravitational lensing leads to distortions in the observed image of distant objects. The deflection angle $\theta^i(\hat n)$ is given by~\cite{Schmidt:2012ne}
\bea
\label{eq:deflection-field-integral}
&& \theta^i(\hat n) = \Pi^i{}_j \left\{ \frac12 n^k (h_o)^j{}_k - B^j_o - \int^{\chi_s}_0 d\chi \left[\left(1-\frac{\chi}{\chi_e}\right) \right.\right. \nn\\
&& \hspace{-0.5cm} \left.\left. \left( \partial^j A + n^k \partial_j B_k - \frac12 n^k n^l \partial^j h_{kl} \right) - \frac{B^j - n^k h^j{}_k }{\chi_e} \right] \right\},
\eea
where $\Pi^i{}_j=\delta^i_j - n^i n_j$. The deflection angle is conventionally decomposed into a gradient potential $\phi(\hat n)$ and a curl potential $\Omega(\hat n)$ through $\theta_i=\nabla_i \phi - \varepsilon_{i}{}^{jk} n_j \nabla_k \Omega$, where $\nabla_i$ is the angular gradient on the two-dimensional sky. In particular, the curl potential, related to the antisymmetric part of the shear tensor $\nabla_i \theta_j$, describes rotation of the image~\cite{Stebbins:1996wx}. Using Eqs.~(\ref{eq:rotation-angle-integral}) and (\ref{eq:deflection-field-integral}), we find that the rotation angle $\psi$ for polarization is related to the angular Laplacian of the curl potential,
\bea
\label{eq:rotation-angle-curl-potential}
\psi(\hat n) = (1/2) \varepsilon_{ij}{}^k n^j \nabla_k \theta^i(n) = - (1/2) \nabla^2 \Omega(\hat n),
\eea
for general linear metric perturbations. The rotation of the photon polarization and the rotation of a lensed image are distinct physical phenomena; while the former is meaningful for an individual light ray, the latter applies to a small but extended object on the sky. However, \refeq{rotation-angle-curl-potential} establishes that the two rotations are quantitatively identified with each other, given that the rotation angle of lensed image is also $\psi_I=-(1/2)\nabla^2 \Omega$~\cite{Dodelson:2003bv,Cooray:2005hm}. This reflects the universal influence of metric perturbations on a bundle of light rays along the geodesic.

A statistically isotropic lens field between a source surface at certain redshift and the observer is characterized by angular (cross-)power spectra for deflection potentials $\phi$, $\Omega$, and the rotation $\psi$. Without other physical mechnisms (e.g. Faraday rotation in a magnetic field) to rotate the polarization, $\psi$ and $\Omega$ are maximally correlated as in \refeq{rotation-angle-curl-potential}, and the (cross-)power spectra are related by
\bea
\label{eq:power-spectrum-relation}
C^{\psi\psi}_J = \left[ J(J+1)/2 \right] C^{\Omega\psi}_J = \left[ J(J+1)/2 \right]^2 C^{\Omega\Omega}_J,
\eea
where $J$ describes the angular scale of the lens. On the other hand, the gradient potential $\phi$, which is a true scalar, does not correlate with the pseudo-scalars $\Omega$ and $\psi$ if the stochastic lens foreground preserves parity.

The effects of weak lensing on the full-sky CMB (cross-)power spectra of both temperature and polarization have been presented in Ref.~\cite{Hu:2000ee} for a gradient deflection field, and in Ref.~\cite{Li:2006si} for both gradient and curl deflection potentials. However, in addition to deflection, rotation of polarization mixes the spin-$2$ Stokes parameters through
\bea
Q(\hat n)\pm i U(\hat n) \longrightarrow e^{\mp 2i\psi(\hat n)} \left[ Q(\hat n)\pm i U(\hat n) \right],
\eea
and therefore modifies the $E$-/$B$-mode multipoles. The mixing takes a form of the direction-dependent cosmic birefringence originally considered for new physics coupled to electromagnetism~\cite{DirectionalBirefringence}. Here, we account for how rotation of the polarization affects $E$-/$B$-mode power spectra, and therefore for the first time present complete results for the lensed CMB power spectra from general linear metric perturbations.

We denote the unlensed CMB power spectra by $C^{XY}_{\ell}$ for $X,Y=\Theta,E,B$, and those after lensing by $\tilde C^{XY}_{\ell}$ with the tilde mark. The lensed power spectra are obtained by averaging over all lens realizations, and therefore preserve statistical isotropy and parity if the lens field does so. The corrections $\delta C^{XY}_{\ell}=\tilde C^{XY}_{\ell}-C^{XY}_{\ell}$, up to linear order in lens power spectra, are (the results presented below have corrected sign mistakes in Ref.~\cite{Li:2006si}, as pointed out in Refs.~\cite{Rotti:2011aa,Padmanabhan:2013xfa})
\begin{widetext}
\bea
\label{eq:deltaCl-ThetaTheta}
\delta C^{\Theta\Theta}_{\ell} & = & - \ell (\ell+1) R C^{\Theta\Theta}_{\ell} + \frac{1}{2\ell+1} \sum_{\ell'J} \left[ C^{\phi\phi}_J \left( F^{\phi}_{\ell\ell'J} \right)^2 P^+_{\ell\ell'J} C^{\Theta\Theta}_{\ell'} + C^{\Omega\Omega}_J \left( F^{\Omega}_{\ell\ell'J} \right)^2 P^-_{\ell\ell'J} C^{\Theta\Theta}_{\ell'} \right], \\
\delta C^{\Theta E}_\ell & = & - \left( \ell^2 + \ell -2 \right) R C^{\Theta E}_\ell - \frac{1}{2\ell+1} \sum_{\ell'J} \left[ C^{\phi\phi}_J F^{\phi}_{\ell\ell'J} G^{\phi}_{\ell\ell'J}  P^+_{\ell\ell'J} C^{\Theta E}_{\ell'} + C^{\Omega\Omega}_J F^{\Omega}_{\ell\ell'J} G^{\Omega}_{\ell\ell'J} P^-_{\ell\ell'J} C^{\Theta E}_{\ell'} \right] \vs
&& + \frac{2}{2\ell+1} \sum_{\ell'J} \frac{2}{J(J+1)} C^{\Omega\psi}_J F^{\Omega}_{\ell\ell'J} H^{\psi}_{\ell\ell'J} P^-_{\ell\ell'J} C^{\Theta E}_{\ell'}, \\
\delta C^{EE}_\ell & = & - \left( \ell^2 + \ell -4 \right) R C^{EE}_\ell - 4 S C^{EE}_\ell \nn\\
&& + \frac{1}{2\ell+1} \sum_{\ell'J} \left[ C^{\phi\phi}_J \left( G^{\phi}_{\ell\ell'J} \right)^2 \left( P^+_{\ell\ell'J} C^{EE}_{\ell'} + P^-_{\ell\ell'J} C^{BB}_{\ell'} \right) + C^{\Omega\Omega}_J \left( G^{\Omega}_{\ell\ell'J} \right)^2 \left( P^-_{\ell\ell'J} C^{EE}_{\ell'} + P^+_{\ell\ell'J} C^{BB}_{\ell'} \right) \right]  \nn\\
&& + \frac{4}{2\ell+1} \sum_{\ell'J} \left[ \frac{4}{J^2(J+1)^2} C^{\psi\psi}_J \left( H^{\psi}_{\ell\ell'J} \right)^2 - \frac{2}{J(J+1)} C^{\Omega\psi}_J  G^{\Omega}_{\ell\ell'J} H^{\psi}_{\ell\ell'J} \right] \left( P^-_{\ell\ell'J} C^{EE}_{\ell'} + P^+_{\ell\ell'J} C^{BB}_{\ell'} \right), \\
\label{eq:deltaCl-BB}
\delta C^{BB}_\ell & = & - \left( \ell^2 + \ell -4 \right) R C^{BB}_\ell - 4 S C^{BB}_\ell \nn\\
&& + \frac{1}{2\ell+1} \sum_{\ell'J} \left[ C^{\phi\phi}_J \left( G^{\phi}_{\ell\ell'J} \right)^2 \left( P^-_{\ell\ell'J} C^{EE}_{\ell'} + P^+_{\ell\ell'J} C^{BB}_{\ell'} \right) + C^{\Omega\Omega}_J \left( G^{\Omega}_{\ell\ell'J} \right)^2 \left( P^+_{\ell\ell'J} C^{EE}_{\ell'} + P^-_{\ell\ell'J} C^{BB}_{\ell'} \right) \right]  \nn\\
&& + \frac{4}{2\ell+1} \sum_{\ell'J} \left[ \frac{4 }{J^2(J+1)^2} C^{\psi\psi}_J \left( H^{\psi}_{\ell\ell'J} \right)^2 - \frac{2}{J(J+1)} C^{\Omega\psi}_J  G^{\Omega}_{\ell\ell'J} H^{\psi}_{\ell\ell'J} \right] \left( P^+_{\ell\ell'J} C^{EE}_{\ell'} + P^-_{\ell\ell'J} C^{BB}_{\ell'} \right),
\eea
\end{widetext}
where we define the {\it mean square} deflection angle $R=\sum_J [J(J+1)(2J+1)/(8\pi)](C^{\phi\phi}_J+C^{\Omega\Omega}_J)$, and the {\it mean square} rotation angle $S=\sum_J [(2J+1)/(4\pi)]C^{\psi\psi}_J$. The five lensing kernels involved are given explicitly by
\begin{widetext}
\bea
F^{\phi}_{\ell\ell'J} & = & F^{\Omega}_{\ell\ell'J} = - \sqrt{J(J+1)\ell'(\ell'+1)} \sqrt{\frac{\Pi_{\ell\ell'J}}{4\pi}} \wignertj{\ell}{0}{J}{-1}{\ell'}{1}, \\
\label{eq:kernel-Gphi}
G^{\phi/\Omega}_{\ell\ell'J} & = & \sqrt{\frac{J(J+1)}{2}} \sqrt{\frac{\Pi_{\ell\ell'J}}{4\pi}} \left[ \sqrt{\frac{(\ell'+2)(\ell'-1)}{2}} \wignertj{\ell}{2}{J}{-1}{\ell'}{-1} \pm \sqrt{\frac{(\ell'+3)(\ell'-2)}{2}} \wignertj{\ell}{2}{J}{1}{\ell'}{-3} \right], \\
\label{eq:kernel-H}
H^{\psi}_{\ell\ell'J} & = & \frac{J(J+1)}{2} \sqrt{\frac{\Pi_{\ell\ell'J}}{4\pi}} \wignertj{\ell}{2}{J}{0}{\ell'}{-2},
\eea
\end{widetext}
where we have introduced the short-hand notation $\Pi_{\ell_1 \ell_2 \cdots}\equiv(2\ell_1+1)(2\ell_2+1)\cdots$. These results consistently satisfy {\it power conservation}; i.e., lensing conserves both $\sum_\ell (2\ell+1)C^{\Theta\Theta}_\ell$ and $\sum_\ell(2\ell+1)\left[C^{EE}_\ell+C^{BB}_\ell\right]$ for arbitrary intrinsic CMB power spectra and arbitrary lens power spectra, because deflection and rotation both redistribute power but do not create anisotropies.

\begin{figure}[ht!]
\centering
\hspace{-0.5cm}
\includegraphics[scale=0.95]{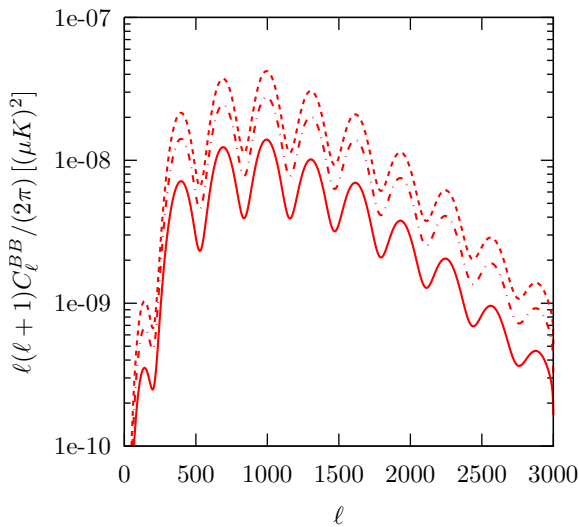}
\caption{(color online). Lensing $B$-mode power spectrum from inflationary gravitational waves with $r=0.13$. We compare contributions from deflection (dashed), from rotation (dash-dotted), as well as the full result including deflection-rotation cross-correlation (solid).}
\label{fig:Bmode}
\end{figure}

\begin{figure}[ht!]
\centering
\hspace{-0.5cm}
\includegraphics[scale=0.95]{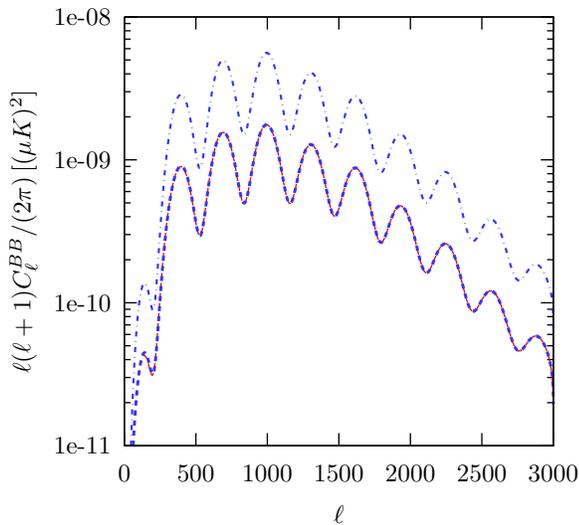}
\caption{(color online). We artificially set $C^{\Omega\Omega}_J=C^{\phi\phi}_J$ and compare the efficiencies with which lensing converts $E$ mode into $B$ mode by $\phi$ only (red, thin dashed), by $\Omega$ without $\psi$ (blue dash-dotted), and by $\Omega$ with $\psi$ (blue, thick dashed).}
\label{fig:efficiency}
\end{figure}

Parity conservation imposes a selection rule on $\ell+\ell'+J=$~even or odd through $P^{\pm}_{\ell\ell'J}=(1+(-1)^{\ell+\ell'+J})/2$. For example, lensing can generate $B$-mode power (on scale $\ell$) from $E$-mode power (on scale $\ell'$) through curl-type deflection $\Omega$ (on scale $J$); since $E$-mode polarization is parity-even, while $B$-mode polarization and $\Omega$ are parity-odd, the coupling exists only if $\ell+\ell'+J=$~even.  

Rotation of polarization modifies previous lensing results through the rotation-angle power spectrum and its cross-correlation with the curl deflection. Therefore, In \refeqs{deltaCl-ThetaTheta}{deltaCl-BB}, terms either proportional to $S$ or involving $C^{\psi\psi}_J$ and $C^{\Omega\psi}_J$ are new. In particular, the new effects modify the prediction for $B$-mode power converted from $E$-mode power by lensing.

\begin{figure}[ht!]
\centering
\includegraphics[scale=0.5]{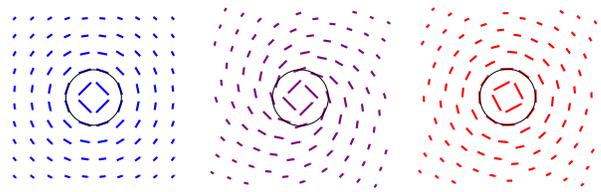}
\caption{(color online). An unlensed polarization map with only $E$ modes (left) is deflected by a curl potential $\Omega$, which is chosen to give uniform $\psi$. The black circle illustrates that pixels are shifted along the tangential direction (clockwise) of concentric circles. Deflection generates a map with an apparent $B$-mode component (middle), but the $B$ modes are reduced once rotation is included (right).}
\label{fig:illustrate}
\end{figure}

Although the results derived above are entirely applicable to general metric perturbations, we illustrate for a stochastic background of gravitational waves from inflation. Assuming the WMAP+BAO+$H_0$ best-fit cosmology~\cite{Komatsu:2010fb}, we consider a scale-invariant primordial tensor power spectrum with tensor-to-scalar ratio $r=0.13$. In \reffig{Bmode} where the power spectrum for the lensing-induced CMB polarization $B$ modes is plotted, it can be seen that the effects of deflection and rotation cancel each other due to cross-correlation, yielding a prediction for these polarization $B$ modes that is $\sim 4$ times smaller than what one would anticipate from deflection alone~\cite{Rotti:2011aa,Padmanabhan:2013xfa}. In \reffig{efficiency}, we artificially set $C^{\Omega\Omega}_J=C^{\phi\phi}_J$ ($C^{\psi\psi}_J$ and $C^{\Omega\psi}$ then follow from \refeq{power-spectrum-relation}). If the rotation $\psi$ is neglected, we find agreement with Ref.~\cite{Padmanabhan:2013xfa} that $\Omega$ is more efficient than $\phi$ at converting $E$ modes into $B$ modes. It is striking, however, that including the rotation coherently with $\Omega$ yields {\it exactly} the same $B$-mode power as induced by $\phi$, as is evident from the coincidence of the two dashed curves in the plot.

Why should deflection and rotation cancel instead of enhance each other in $B$-mode generation? \reffig{illustrate} provides a heuristic picture, where we start with a polarization map of concentric patterns with only $E$-mode (left panel). A curl potential, which we choose to give uniform $\psi$ via \refeq{rotation-angle-curl-potential}, shifts pixels in the tangential direction along concentric circles, and therefore induces a $B$-mode component (middle panel). The curl deflection therefore effectively ``rotates'' the sky about the origin, but keeps the polarization orientation fixed. However, the gravitational field also rotates the polarizations (right panel), in particular with an angle $\psi$ exactly equal to the rotation angle of the patch. This to large extent undoes the spurious $B$-mode generated by deflection. If $\psi$ were to rotate polarizations in the opposite way the sky rotates, this cancellation would not occur.

Lensing $B$ modes from $\Omega$ and from $\psi$ should not be regarded as separate signatures, since metric perturbations always generate $\Omega$ and $\psi$ simultaneously and coherently. Therefore, our results can be better interpreted as necessary corrections to the previous (incomplete) results for curl-deflection of polarizations. Reconstruction schemes for $\Omega$, previously developed to either monitor systematics~\cite{Cooray:2005hm} or to constrain primordial vector/tensor perturbations~\cite{Namikawa:2011cs}, have left out the contribution from rotation. Methods to measure direction-dependent $\psi$~\cite{DirectionalBirefringence}, on the other hand, cannot be directly applied to the case of lensing, because correlation with $\Omega$ is not included. Reconstruction schemes should combine the effects from $\Omega$ and that from $\psi$.

We emphasize that the leading lensing signature that is pursued by current observations, i.e. that from linear scalar perturbations due to large-scale structure, is unaffected by our more general results. Still, beyond linear order in perturbation theory, weak lensing by large-scale structure is expected to rotate the polarization. Large-scale vorticity flow can contribute a signal by generating a vector mode in the metric. Moreover, rotation can appear at second order in potential perturbations, since it is related to the image rotation which arises from lens-lens coupling~\cite{Cooray:2002mj,Hirata:2003ka,Cooray:2005hm,Pen:2005fy}. Rotation will thus need to be included when the weak lensing of polarization has to be studied at second order in scalar perturbations.

Our discussion may apply to other observations in cosmology and astrophysics, e.g. future 21-cm surveys or strongly lensed quasars, where lensing distortions to the observed photon polarization might need to be studied.


\begin{acknowledgments}

The author would like to thank Marc Kamionkowski, Aditya Rotti, Christopher Hirata for useful discussions and comments. This work is supported by the William Gardner Fellowship, and by the National Science Foundation under Grant No. PHYS-1066293.

\end{acknowledgments}


\end{document}